# Improving Overhead Computation and pre-processing Times for Grid Scheduling System

[1]Asgarali Bouyer, [2]Mohammad javad Hoseyni,
Department of Computer Science
[1,2]Islamic Azad University-Miyandoab branch
Miyandoab, Iran
[1]basgarali2@live.utm.my, smj.hosseini@gmail.com

Abdul Hanan Abdullah
Faculty of Computer Science and Information Systems
UNIVERSITI TEKNOLOGI MALAYSIA
Johor, Malaysia
hanan@utm.my

*Abstract*— Computational Grid is enormous environments with heterogeneous resources and stable infrastructures among other Internet-based computing systems. However, the managing of resources in such systems has its special problems. Scheduler systems need to get last information about participant nodes from information centers for the purpose of firmly job scheduling. In this paper, we focus on online updating resource information centers with processed and provided data based on the assumed hierarchical model. A hybrid knowledge extraction method has been used to classifying grid nodes based on prediction of jobs' features. An affirmative point of this research is that scheduler systems don't waste extra time for getting up-to-date information of grid nodes. The experimental result shows the advantages of our approach compared to other conservative methods, especially due to its ability to predict the behavior of nodes based on comprehensive data tables on each node.

*Keywords-component; job scheduling; hierarchical model; Grid nodes modul; Grid resource information center*

## I. INTRODUCTION

In computational grid systems, a job or application can be divided into tasks and distributed to grid nodes. These tasks can be executed independently at the same time in parallel ways to minimize completion time of job execution. Therefore, grid nodes dynamically share their resources to use by another application. In order to perform job scheduling and resource management at Grid level, usually it is used a meta-scheduler. A resource scheduler is fundamental in any large-scale Grid environment. The task of a Grid resource broker and scheduler dynamically to identify and characterize the available resources, and to select and allocate the most appropriate resources for a given job. In a broker-based management system, brokers are responsible for selecting best nods, ensuring the trustworthiness of the service provider. Resource selection is an important issue in a grid environment where a consumer and a service provider are distributed geographically across multiple administrative domains. Choosing the suitable resource for a user job to meet predefined constraints such as deadline, speedup and cost of execution is an main problem in grids. As you know, each task has some conditions that must be considered by schedulers to select the destination nodes based on the place of tasks or applications. For example, suitable node selection can reduce overhead communication and cost and makespan and even execution time. Resource discovery is important but not enough because of the dynamic variation in the grid, such that resource prediction is necessary for grid system to predict coming status of nodes and their workloads. Therefore, for prediction of node's status, schedulers need to get up-to date or last information about nodes. Another problem is how to get up-to date information about nodes. In most of the grid systems, there are some special centers that maintain last information about grid node's status that periodically updated by its management section such as Meta-computing Directory Services [1] in Globus toolkit. In the Globus Toolkit, Resource and status information is provided via a LDAP-based network directory called Meta-computing Directory Services (MDS). It has a grid information service (GIS) that is responsible for collecting and predicting the resource status information, such as CPU capacities, memory size, network bandwidth, software availabilities, and load of a site in a particular period. GIS can answer queries for resource information or push information subscribers [2]. n our research, we have used GIS idea to maintain nodes' information, but a little different from Globus' GIS, for predicting in a local fashion. For this aim, we used a special component on all participant Grid nodes that is called grid node's module (GNM). In Globus, all processing information is done by MMDS, and it does not use local processing for this purpose. However, we have used a local information center each node to maintain a complete information or background about its status n order to exactly exploration of knowledge for valuation and scheduling.

The rest of this paper is ordered as follow. In section two, a problem formulation is described. Some, related works on earlier research have been reviewed in section 3. Our proposed approach has been discussed in section4. In section 5, the experimental results and evaluations have been mentioned. Finally, the paper is concluded in section 6.

## II. PROBLEM FORMULATION

One motivation of Grid computing is to aggregate the power of widely distributed resources, and provide non-trivial services to users. To achieve this goal, an efficient Grid scheduling system must be considered as an essential part of the Grid. Since the grid is a dynamic environment, the prediction and detection of available resources and then use an economic policy in resource scheduling for coming jobs with consider some sensible criteria is important in scheduling cycle. In a Grid environment, prediction of resource availability, allocation of proper nodes to desired tasks, a fairly price adapter for participant nodes is the prerequisite for a reasonable scheduling guarantee. Many approaches for grid meta-scheduler are discussed from different points of view, such as static and dynamic policies, objective functions, application models, adaptation, QOS constraints, and strategies dealing with dynamic behavior of resources that have some weaknesses (e.g., complexity time, predicting problems, using out of date data, unfair, inflexible, nonflexible, etc.). Based on the current researches, a new approach has been proposed as a helpful tool for meta-scheduler to do a dynamic and intelligent resource scheduling for grid with considering some important criteria such as dynamism, fairness, response time, and reliability.

The job scheduling problem is defined as the process of making decision for scheduling tasks of job based on grid resources and services. Grid scheduling problem is formally represented by a set of the given tasks and resources. A grid system is composed of a set on nodes as $N = \{N_1, N_2, ..., N_n\}$ and each node consists of several resources, that is, $N_i = \{R_1^i, R_2^i, ..., R_r^i\}$ and each resource is appeared often in all nodes within different characteristics. By a set of the given jobs in time period T, it consists of several jobs within different characteristics, that is, $J = \{J_1, J_2, ..., J_j\}$ that belong to c consumers $C = \{C_1, C_2, ..., C_c\}$. Each job necessarily is divided into several tasks, that is, $J_i = \{T_1^i, T_2^i, T_3^i, ..., T_t^i\}$. The main objective in most scheduling systems often is to design a scheduling policy for scheduling submitted jobs with the goal of maximizing throughput and efficiency and also minimizing job completion times. Job's scheduling is generally broken down into three steps:

1- To define a comprehensive and versatile method and divide fairly job between grid nodes.
2- The allocation of tasks to the computing nodes based on user requirement and grid facilities.
3- The monitoring of running grid tasks on the nodes over time and reliability factors.

With a large number of users attempting to execute jobs concurrently on the grid computing, parallelism of the applications and their respective computational and storage requirements are all issues that make the resource scheduling problem difficult in these systems.

## III. RELATED WORKS

Condor's Matchmaker [3-5] adopts a centralized mechanism to match the advertisement between resource requesters and resource providers. However, these centralized servers can become bottlenecks and points of failures. So the system would not scale well when the number of the nodes increases.

AppLeS (Application Level Scheduling) [6] focuses on developing scheduling agents for individual Grid applications. It applies agents for individual Grid applications. These agents use application oriented scheduling, and select a set of resources taking into consideration application and resource information. AppLeS is more suitable for Grid environment with its sophisticated NWS[7] mechanism for collecting system information [8]. However, it performs resource discovering and scheduling without considering resource owner policies. AppLeS do not have powerful resource managers that can negotiate with applications to balance the interests of different applications [8]. EMPEROR [9] provides a framework for implementing scheduling algorithms based on performance criteria. The implementation is based on the Open Grid Services Architecture (OGSA) and makes use of common Globus services for tasks such as monitoring, discovery, and job execution. EMPEROR is focused on resource performance prediction and is not distributed nor does it support economic allocation mechanisms.

Singh et al. proposed an approach for solving the Grid resource management problem by taking into consideration[10]. The paper proposed an approach aimed at obtaining guarantees on the allocation of resources to task graph structured applications. In mentioned research, resource availabilities are advertised as priced time slots, and the authors presented the design of a resource scheduler that generates and advertises the time slots. Moreover, Singh et al. demonstrated that their proposed framework (incorporating resource reservation) can deliver better performance for applications than the best effort approach.

Another work has been done by Chao et al. that is a coordination mechanism based on group selections of self-organizing agents operating in a computational Grid [18]. The authors argued that due to the scale and dynamicity of computational Grids, the availability of resources and their varying characteristics, manual management of Grid resources is a complex task, and automated and adaptive resource management using self-organizing agents is a possible solution to address this problem. Authors have pointed out that for Grid resource management, examples in which performance enhancement can be achieved through agent-based coordination include: decision making in resource allocation and job scheduling, and policy coordination in virtual organizations.

Kertész and Kacsuk have argued that there are three possible levels of interaction to achieve interoperability in Grids: operating system level, Grid middleware level, and higher-level services level [11]. Authors described three approaches to address the issue of Grid interoperability, namely: 1) extending current Grid resource management systems; 2) multi-broker utilization; and 3) meta-brokering. In extending current Grid resource management systems, they developed a tool called GTBroker that interacts with Globus resources and performs job submission. This proposed meta-brokering service is designed for determining which Grid broker should be best selected and concealing the differences in utilizing them. Extra to the meta-brokering service, they proposed a development of the Broker Property Description Language (BPDL) that is designed for expressing metadata about brokers. The authors have also implemented their ideas in the Grid meta-broker architecture that enables users to access resources of different Grids through their own broker(s).

Many other considerable approach such as hierarchical grid resource management [12], a new prediction-based method for dynamic resource provisioning and scaling of MMOGs in grid [13], aggregated resource information for resource selection methods by grid broker[14] has been offered with considerable idea that is recommended for researches as hopeful methods.

## IV. GRID NODE'S MODULE FOR OPTIMIZED SCHEDULING

Most of grid scheduling systems consist of two main components: nodes, and schedulers. Scheduler can be considered as local schedulers and meta-schedulers. In some earlier methods [3, 15-18] meta-scheduler, as the main component, are responsible for job scheduling. However, there is other scheduling methods [19-23] in which local schedulers perform most of job scheduling steps. These mentioned methods have not applied the impact of using grid nodes in scheduling system and they only map jobs to nodes. In this section we are going to devolve some steps of scheduling process to grid nodes or all participant nodes in grid system.

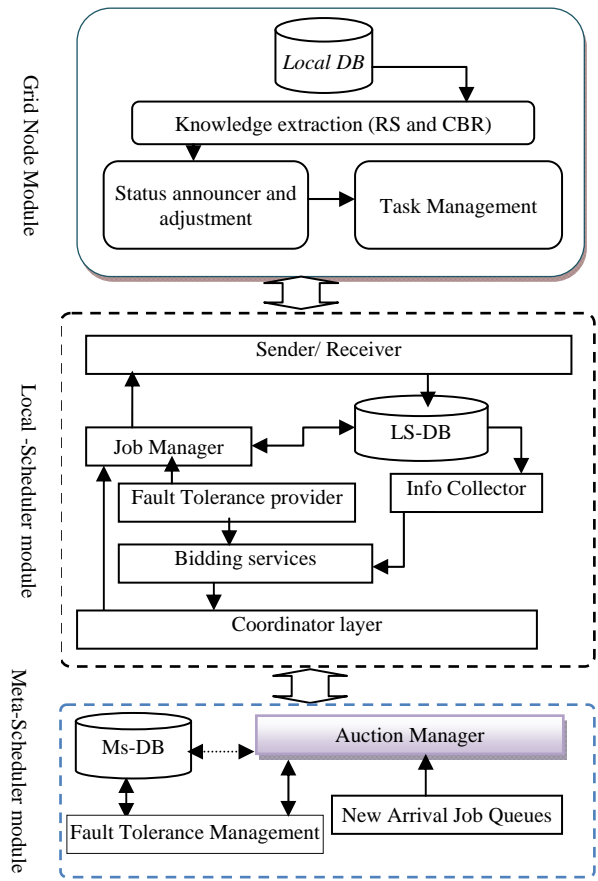

Figure 1. A hierarchical architecture for optimized scheduling.

A general architecture of the grid scheduling system has been depicted in Fig.1. Since, this architecture uses an auction mechanism by meta-scheduler and participant local schedulers for job submission and resource allocation like other methods, we only focus on Grid Node's Module (GNM) as a significant part of our research. Note that the model described here does not prescribe any implementation details; the protocols, programming languages, operating systems, user interfaces and other components. Proposed architecture uses a hierarchical model with minimum communication cost and time.

In this research, the knowledge extraction module is devolved to Provider Node (PN). In many approaches [24], the needed information is gathered in special places in order to manage by Grid Resource Brokers or Meta-Schedulers that surely take much time or have the problem of out-of-date information. Here, the proposed module for provider node saves all required information in the local database and it will do knowledge extraction methods in a local fashion. Finally, the summarized information about each grid node's status is saved in local scheduler's data tables and dynamically is updated by an online method [25]. A new illustration of GNM is depicted in Fig. 2 with more details.

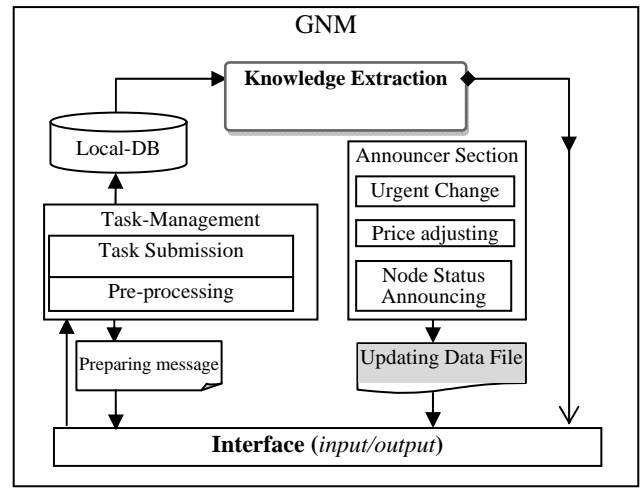

Figure 2. The Grid Node's Module (GNM) with more details.

### A. Knowledge Extraction

The applied methods for knowledge extraction are Rough Set theory and Case-based Reasoning. GNM uses Case-Based Reasoning (CBR) technique and Rough Set Analysis. These techniques work together to discover knowledge in order to supply a satisfied recommendation to send to local scheduler. The exploration is based on previous data or experiments on the "node data table" that is helpful to make a better decision. Use of multiple knowledge extraction methods is beneficial because the strengths of individual technique can be leveraged. In previous research [18], we proposed a learning method for resource allocation based on the fuzzy decision tree. This method observed that it has a successful potential to increase accuracy and reliability if the job length is large. However, in this section, we use a hybrid of CBR and RS to get the exact knowledge with considering economic aspects. This section is divided in three sub-sections: Rough Set Analyzer, Case-based reasoning method, and calculating some information for computing of priority.

We used Rough Set (RS) theory [26] to generate rules in order to analyze by GNM to classify proper nodes to use by CBR method. Rough set analysis provides an effective means for analysis of data by synthesizing or constructing approximations (upper and lower) of set concepts from the acquired data. It also proved to be very useful for analysis of decision problems concerning objects described in a data table by a set of condition attributes and decision attributes. The goal of using the rough set theory in this research is to generate useful rules for classifying similar states to apply by CBR in order to explain the best state for node to accept or reject this new task. Our purpose is to carry out resource selection for the desired job based on job condition in the scheduling phase. To do this issue, we will use Rough Set Analyzer (RSA) to generate rules. It takes the nodes' information data table as input. The output is three Matrixes (generated rules are shown in matrix form).

The RSA uses three important attributes (final status of task, completion time, and cost price) as decision attributes. These attributes can be acted upon as the condition attributes and decision attribute of a decision system. Desired application only uses one of this attributes at a moment as decision attribute and at the same time, other two attributes will be considered as conditional attributes. For example, if dependability and speed factors are more important, the second and third attribute is considered as Decision attribute, respectively. There are other conditional attributes that we have mentioned in next section. In addition, RSA needs to discretize the input data for some attributes. Since RSA takes analysis time in order to perform the rough set method, though not considerable, it is possible that we are encountered with this question: When will RSA execute rough set analysis? To answer this question, we supply two conditions for doing a rough set analysis:

Number of currently added tasks to this node is more than 1% of previous submitted tasks in the past days.

Rough set analysis has not been done in last 24 hours.

Case-based Reasoning is a technique that adapts past solutions for new demands by using earlier cases to explain, criticize and interpret novel situations for a new problem [27]. The basic CBR processes are defined as a cycle and include the following: RETRIEVE the most similar cases; REUSE the cases to solve the problem; REVISE the proposed problem solution; RETAIN the modified solution as a new case. These steps fully must be done to get the satisfied knowledge. Now, we encounter with this question: When will Case-based Reasoning be executed? For this question, first we should say that when will the nodes get the new tasks (or job) information? During online resource selection by local scheduler, the job information is sent to all nodes. In [28] an optimized version of Case-Based Reasoning had been proposed to increase accuracy in final results. This method applies CBR algorithm by using Decision Tree in order to select suitable sampling. Improving accuracy criterion was a success key in this method. However, due to classification of input data by data mining techniques such as decision tree, selecting training set takes much time that is not negligible for online processes. Therefore, to reduce of this overhead time, we use rough set rules to classify and define training set. It consists of two steps: 1) selecting consistent rules for the job in order to get desired samples (records) to define training sets. In this case, it can select a best training set. 2) Final processing and predicting the situation of a coming job by using neighboring records (in the same class).

After doing CBR, the obtained knowledge about job and job (executing job on this node) will be sent to scheduler. In the next sections, we will describe how local scheduler use this extracted knowledge for resource allocation.

### B. Task Management

Since the capacity of resources in each node is changed at the moment, new task must be processed before submitting because the existing capacity may not be sufficient for a desired task in determined deadline. In this case, task is not inserted to queue and rejection information is sent to local scheduler (LS). This operation is done after CBR execution and the result is sent along with extracted knowledge (by CBR). In contrast, if the existing resources be enough for this task, it will be successfully submitted in the queue of the grid's task on the provider node. All information about this task is inserted in the related data table as a new *Recordset*. GNM record several important properties at this time such as CPU Load, Free memory (RAM), Task-ID, size of the new task, priority of the new task (we consider only 3 priority Low, Normal and High), number of all grid tasks (in waiting status), amounts of Data Transmission Rate (DTR) related to this node in the grid (DTR probably has upheaval in sometimes), start time of task execution, spent time for this task, completion time, status of a task (wait, running, success, and fail). Some of this information (e.g. spent and completion time, task status and so on) is updated after finishing a task. In our approach, task has four states: wait, running, fail, and success. After submit a task in the queue, at first, it take wait state. When a task is started for executing, its state changes to running state until it is terminated. After successfully finishing, the task state will be changed to success state. It is possible that task state is changed to fail state due to diverse software and hardware problems. At the end, the result completely is given back to LS if the task successfully is executed.

### C. Announcer Section

This section is the most important section in GNM. It is responsible to decide on whether the node is ready to accept a new task or deny new task. Announcer section (AS) analyzes the grid tasks queue and its own status (mentioned in above) to specify coming status. For example, it specifies that in next two hours it cannot accept any new task. This section is definitely analyzing its own status after every submitting. It evaluates deadline and execution time of the previous submitted task (waiting and running state) to determine how many processes in the near future will finish. With the assumption of finishing these processes, when would the desired node be able to accept new tasks in the future? In addition, it is possible that some high priorities local processes will join to current processes in near future (e.g. automatically start a Virus Scan program, Auto saves or backup by some application, and so forth). Thus, AS has to consider all possible status to get the best decision. This process will be done by sub-section that is called Node Status Announcing (NSA) module.

NSA module also computes some related information, such as Success Ratio, Average of Completion Time (ACT), Average of CPU-Idle (how much percent is CPU free or idle) and Average of free memory (RAM), about this node and sending it along with other obtained results to Local scheduler. For instance, ACT measure is computed as following equation:

$Success\ Ratio = N_s/N_a$

$$ACT_k = (\sum_{i=1}^{n} GTp_i)/n \qquad (1)$$

$GTp_i$ is completion time for task i; and n is the number of success tasks by node $k^{th}$.

$N_s$: Number of successfully completed tasks.
$N_a$: Number of Successful + Failed tasks

It is mentioned that aborted tasks are different from failed task. Fail event can be occurred because of a nodes' problem such as software, hardware, deadline or budget problems. Where abort event is done by scheduler for that canceling of job by consumer or other problems and executive node has not any problem for continuing job execution. Therefore, aborted task is considered as neutral tasks and those are not taken into account for measuring of the success ratio.

Sometimes a node is encountered with unpredictable cases. For example, suppose that a desired node is ready to accept new tasks. If node's resources have unexpectedly been occupied by local tasks (OS processes), this node cannot accept a new task until to come back to normal state. In this case, Urgent Change section, a sub-section in Announcer Section, has to change its status to non-acceptance and then inform this change to scheduler. After come back to normal state, this section has to announce it to Local Scheduler.

Another subsection is Price adjusting section. This module is responsible for determining the price of a node based on standard factors and the current node status. For example, if the computed price based on standard parameters for one minute become $\alpha$, this module can change this price based on current status such as the number of submitted tasks (in waiting state), number of success tasks/ number of failure tasks in last day and last week and so on. Its mention that, due to respect for grid owners and grid costumers profits, the price increment or decrement can be in the following range:

$\alpha*(1-p) < $ Offered Price $< \alpha*(1+p)$ : $\alpha$ is standard price, and $0 \le p \le 0.5$

At the end, this Offered Price is sent to local scheduler. Therefore, the offered price by provider node always is dynamic.

## V. EXPERIMENTAL RESULTS AND DISCUSSION

To observe the effect of GNM architecture, we used GridSim simulator [29]. GridSim support hierarchical and economic-based grid scheduling systems and it is used as a reference simulation environment for most of the significant research such as [30-32] and compare our results with the job scheduling algorithms proposed in [18, 33].

Four important measures were evaluated in the simulation: dependability or reliability, accuracy prediction, and success ratio and iteration of job in other nodes. In GridSim, each simulated interactive component (e.g. resource and user) is an entity that must inherit from the class GridSim and override a body()method to implement its desired behavior. Class Input and Output in GridSim is considered for interacting with other entities. Both classes have their own *body*() method to handle incoming and outgoing events, respectively. Entities modeled in GridSim include the resources, users, information services, and network-based I/O.A resource, that in our method called provider node, is characterized by a number of processors, speed of processing (a specialized CPU rate for the grid task), The tolerance of price variation for provider node, The real data transmission rate per second, The capacity of RAM memory, Monetary unit, Tolerance of the price variation and time zone. Furthermore, the node's price is computed based on mentioned characteristics. Tolerance of price variation is a parameter to give a discount over node's price that is used for some low budget jobs. For each resource, the CPU speed has been determined by MIPS measure (million instructions per second). Each property is defined in an object of the *ResourceCharacteristics* class. The flow of information from other entities via I/O port can be processed by overriding *processOtherEvent*() method. We used a uniform allocation method for all nodes [23].

For our testing, we define three different local scheduler (Table 1) and three groups of jobs (Table2). Each group of jobs has special features that have been mentioned in Table 2. In our previous work [18] the nodes' specifications and their performance is collected from a real local grid. In this research, the updated of this data table is used for supposition nodes, and so we do not explain about nodes' properties.

Each group of jobs is submitted on different times all three local schedulers. It is necessary to say that, tasks of jobs are submitted in a parallel form on available nodes in every local scheduler. For example, the *Job_Group1* is composed of 250 tasks and 45500 Million Instructions for every task that each task averagely has 1200 second deadline to complete integrally. Each group of jobs has been tested 15 times separately on each local scheduler's node.

Since, most of the presented scheduling systems and scheduling algorithms were tested and evaluated based on specific assumptions and parameters of authors, therefore, nobody cannot claim that his/her method is the best. However, in this research we tried to test of our approach in GridSim simulator with developing real nodes' behavior for node (resource) entity. The following experiments show the comparison of GNM effect to use in node selection step by local schedulers. In Fig. 3 the number of tasks' completion is compared for new approach and recent work [18].

The analysis of obtained results in Fig. 3 show that, due to use rough set based case base reasoning on grid node module (it is not necessary online), the workload of schedulers is decreased and so the overhead time for selection node is decreased. Consequently, as it is seen in Fig. 3, the overhead time for node selections, starting, and gathering results plus execution time for the new proposed approach is less than the considered deadline for each task of job_Groups3 on local scheduler LS1, whereas this time in fuzzy based scheduling method [18] totally is more than the task deadline. Therefore, any task of job_Groups3 cannot be finished in considered deadline on LS1 based on fuzzy based scheduling method.

However, decision accuracy for the new approach was low rather than other methods for jobs that reliability and completion factors nearly had equal priority because of especial priority computing by each method.

TABLE I. THE DESCRIPTION OF CONSIDERED LOCAL SCHEDULERS.

| Local scheduler's (**LS**) Name | Number of available Nodes | GMIPS (MIPS) allocated CPU MIPS for grid tasks | current queue deadline status (sec) | Medium node's dependability |
|---|---|---|---|---|
| LS1 | 400 | 65 | 460 | 0.72 |
| LS2 | 320 | 140 | 350 | 0.93 |
| LS3 | 750 | 80 | 400 | 0.85 |

TABLE II. SAMPLE OF JOBS.

| Groups of jobs name | Number of jobs | Deadline for each task (sec) | Memory for each task (MB) | Task length by (Million Instruction) (MI) | reliability | completion time |
|---|---|---|---|---|---|---|
| Job_Group1 | 5 job (250 task) | 1200 | 1.93 | 45500 | 0.8 | 0.2 |
| Job_Group2 | 3 job (210 task) | 2100 | 3.4 | 72000 | 0.3 | 0.7 |
| Job_Group3 | 5 job (100 task) | 900 | 6.25 | 30000 | 0.5 | 0.5 |

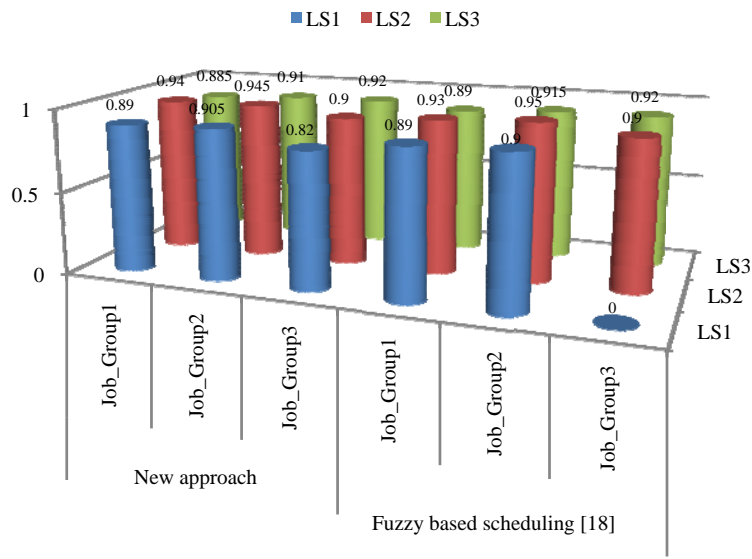

Figure 3. The ration of completed tasks of jobs on three local schedulers for new approach and earlier work [18] based on grouped jobs in Table. 2.

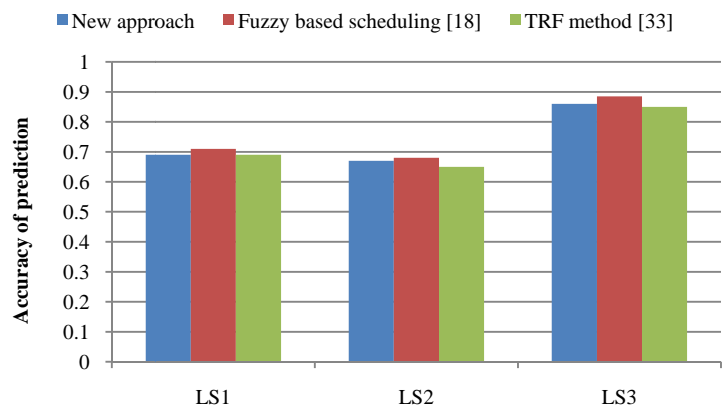

Figure 4. The comparison of accuracy prediction for the new approach and other two methods.

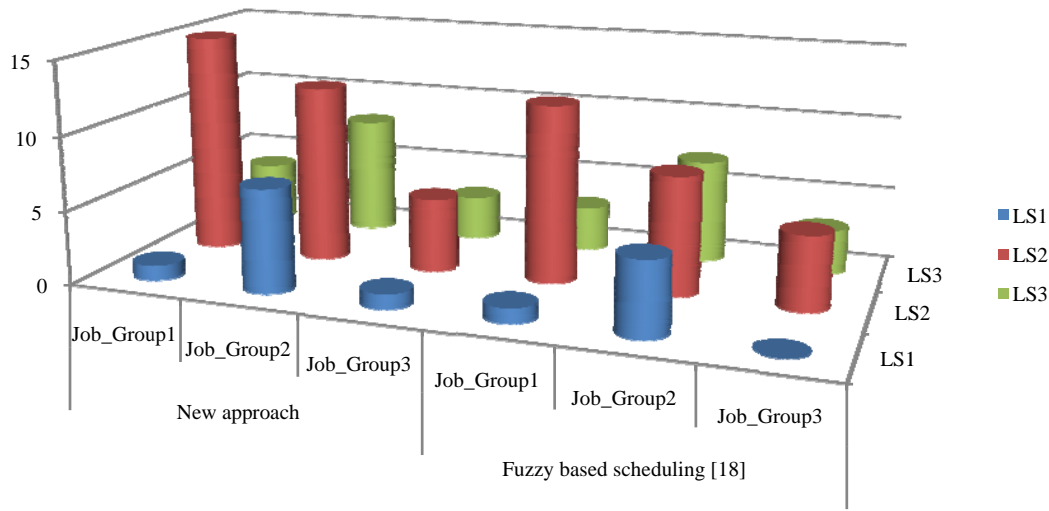

Figure 5. The evaluation of tasks iteration in the new approach and other method.

The evaluation of the Fig. 4 show that the fuzzy based scheduling method has better accuracy prediction rather than new approach and TRF methods [33] because of online decision with up-to date and consuming more time for knowledge extraction. However, TRF method is acceptable when scheduler want to select a small number of nodes between large numbers of available nodes. By the way, overhead time for TRF is less than both other methods and close to our new approach.

The purpose of iteration in this paper is the replace new node instead of faulted node and restart job in new node. Since, the overhead time in fuzzy based method is much, if a task has not been a sufficient deadline, it cannot be iterated on the new node and so this task is failed. According to obtained results in Fig. 5, the new approach has more task iteration than fuzzy scheduling method, and so it will act a bit better than fuzzy-based method based on a number of iterations.

## VI. CONCLUSION

According to the weaknesses of the earlier grid resource discovery approaches, such as using out-of- date information for job scheduling, much elapsed time for task submission, considerable overhead communications and cost, this paper uses an optimal strategy to improve resource discovery efficiency. In this approach, all information about grid nodes is not necessarily aggregated on grid resource information centers. We used a local data table on each grid node to maintain all information about grid node's status. Moreover, grid resource information centers maintain a summary of up-to-date information that continually is updated if there was a significance changing in node's resources or performance. The experimental results explain that our approach reduces the overhead time and improves the resource discovery efficiency in a grid system.